\documentclass[useAMS,usenatbib]{mn2e}
\usepackage{rotating}
\usepackage{journals}

\def\approxgt{\ifmmode \rlap{$>$}{}_{{}_{{}_{\textstyle\sim}}} \else%
$\rlap{$>$}{}_{{}_{{}_{\textstyle\sim}}}$\fi} 
\def\approxlt{\ifmmode \rlap{$<$}{}_{{}_{{}_{\textstyle\sim}}} \else%
$\rlap{$<$}{}_{{}_{{}_{\textstyle\sim}}}$\fi}

\def\farcs{\hbox{$.\!\!^{\prime\prime}$}}

\def\arcsec{\hbox{$^{\prime\prime}$}}

\LARGE \normalsize \title[The companion star of Cir~X--1?]{Detection of the radial velocity
curve of the B5--A0 supergiant companion star of Cir~X--1?}

\author[Jonker, Nelemans, \& Bassa]  {P.G.~Jonker$^{1,2,3}$\thanks{email :
p.jonker@sron.nl. Based on observations made with ESO telescopes at the Paranal Observatories under
programme ID 274.D--5047(A)}, G. Nelemans$^{4}$, C.G.~Bassa$^3$ \\
$^1$SRON, Netherlands Institute for Space Research, Sorbonnelaan 2, 3584~CA, Utrecht, The Netherlands\\
$^2$Harvard--Smithsonian  Center for Astrophysics, 60 Garden Street, Cambridge, MA~02138, Massachusetts,
U.S.A.\\
$^3$Astronomical Institute, Utrecht University, P.O.Box 80000, 3508 TA, Utrecht, The Netherlands\\
$^4$Department of Astrophysics, IMAPP, Radboud University Nijmegen, Toernooiveld 1, 6525 ED, Nijmegen, The Netherlands\\
}

\begin{document}

\maketitle

\begin{abstract} \noindent In this Paper we report on phase resolved $I$--band optical spectroscopic and photometric
observations of Cir~X--1 obtained with the Very Large Telescope. The spectra are dominated by Paschen absorption
lines at nearly all orbital phases except near phase zero (coinciding with the X--ray dip) when the absorption lines
are filled--in by broad Paschen emission lines. The radial velocity curve of the absorption lines corresponds to an
eccentric orbit ($e=0.45$) whose period and time of periastron passage are consistent with the period and phase
predicted by the most recent X--ray dip ephemeris. We found that the $I$--band magnitude decreases from 17.6 to $\sim
16.8$ near phase 0.9--1.0, this brightening coincides in phase with the X--ray dip. Even though it is likely that the
absorption line spectrum is associated with the companion star of Cir~X--1, we cannot exclude the possibility that
the spectrum originates in the accretion disc. However, if the spectrum belongs to the companion star, it must be a
supergiant of spectral type B5--A0. If we assume that the compact object does not move through the companion star at
periastron, the companion star mass is constrained to $\approxlt$10 M$_\odot$ for a 1.4~M$_\odot$ neutron star,
whereas the inclination has to be $\approxgt 13.7^\circ$. Alternatively, the measured absorption lines and their
radial velocity curve can be associated with the accretion disc surrounding a 1.4~M$_\odot$  neutron star and its
motion around the centre of mass. An absorption line spectrum from an accretion disc is typically found when our
line--of--sight passes through the accretion disc rim implying a high inclination. In this scenario the companion
star mass is found to be $\sim$0.4 M$_\odot$. However, from radio observations it was found that the angle between
the line--of--sight and the jet axis is smaller than 5$^\circ$. This would mean that the jet ploughs through the
accretion disc in this scenario, making this solution less probable. 

\end{abstract}

\begin{keywords} stars: individual (Cir~X--1) --- 
accretion: accretion discs --- stars: binaries --- stars: neutron
--- X-rays: binaries
\end{keywords}

\section{Introduction} 

\noindent X--ray binaries are binary systems harbouring a neutron star or black hole compact object that
accretes matter from either a low-- or a high--mass companion star (LMXB and HMXB, respectively).  LMXBs
are typically old systems whereas the early type companion star of HMXBs precludes systems older than a
few times 10$^7$ yr. Recently, it has been realised that many X--ray binaries might have started--off as
intermediate mass X--ray binaries (cf.~Cyg~X--2; \citealt{1999A&A...350..928T};
\citealt{2002ApJ...565.1107P}). One intriguing system that so far has defied classification is Cir~X--1.
It has a 16.6~day orbital period (\citealt{1976ApJ...208L..71K}; see \citealt{2004MNRAS.348..458C} for
the latest X--ray ephemeris). Owing to the detection of type~I X--ray bursts, the compact object in
Cir~X--1 likely is a neutron star (\citealt{1986MNRAS.219..871T}; \citealt{1986MNRAS.221P..27T}). 

The X--ray and radio behaviour of the source is complex. The long--term X--ray lightcurve of the source has been
discussed in detail by \citet{2003HEAD....7.1723S}. One of the striking features of the X--ray lightcurve is the
periodic appearance of dips. In radio, an arc minute scale radio nebula was found (\citealt{1993MNRAS.261..593S}).
Furthermore, a relativistic outflow has been detected on scales of arc seconds that is aligned with the arc minute
scale jet (\citealt{1998ApJ...506L.121F}; \citealt{2004Natur.427..222F}). The inclination of the jet with respect
to the line--of--sight has to be less than 5$^\circ$ (\citealt{2004Natur.427..222F}). The presence of relativistic
outflows detectable as synchrotron emission in the radio band argues against a high magnetic field neutron star in
Cir~X--1 (cf.~\citealt{2000MNRAS.317....1F}). Furthermore, the detection of type~I X--ray bursts from Cir~X--1
suggests that the neutron star has a low magnetic field, since the thermonuclear instability giving rise to  type
I X--ray bursts is suppressed by dipole magnetic fields $\approxgt 10^{12}$ Gauss (\citealt{1980ApJ...238..287J};
\citealt{1995ApJ...438..852B}). In accordance with this, type I X--ray bursts are not found in HMXBs. The low
magnetic field of the neutron star in Cir~X--1 suggests that it is old and hence the companion star is not an
early type star.

So far, clear spectroscopic evidence on the nature of the companion star is lacking. It has been suggested
that Cir~X--1 has a supergiant companion (\citealt{1980A&A....87..292M}). However, it can be argued
(cf.~\citealt{1999MNRAS.308..415J}) that the source is too faint in the optical bands for the distance of
8--10 kpc that has been derived from the type~I X--ray bursts (\citealt{2004MNRAS.354..355J}). Nevertheless,
\citet{2003A&A...400..655C} found emission lines in the near--infrared spectrum that are consistent with a
mid--B super giant. However, these authors concluded that these features must have arisen in the accretion
disc and/or in outflows from the disc.

In this Manuscript we present phase resolved Very Large Telescope (VLT) photometric and spectroscopic observations
obtained with the  FOcal Reducer/low dispersion Spectrograph 2 (FORS2).

\section{Observations, analysis and results} 

We have obtained VLT/FORS2 spectra of the peculiar X--ray binary Cir~X--1 using the 1028z holographic grism
with a slit width of 1\arcsec. The observations were obtained in service mode on 21 different nights in the
period ranging March 15--May 15, 2005 (MJD 53446--53507). In order to sample the $\approx$16.6 day long
binary orbital period of Cir~X--1 we obtained one spectrum per night with an exposure time of 1730 seconds.
Exceptions were May 13 and April 4. On May 13 we obtained 9 spectra with an exposure time of 1675 seconds
since that night Cir~X--1 was close to periastron (according to the ephemeris of
\citealt{2004MNRAS.348..458C}). On April 4, three spectra were obtained since the seeing was higher than the
specified conditions on 2 of the 3 occasions. To minimise the light coming from an unrelated nearby field
star (star 2 in \citealt{1992A&A...260L...7M}), observations were obtained under good seeing conditions
(seeing between 0\farcs4 and 1\farcs3~as measured from the point spread function
full--width--at--half--maximum of the acquisition images). The dispersion was 0.86~\AA~per pixel. With the
1\arcsec~slit width the resolution is about 120 km s$^{-1}$ at 8800~\AA.

The spectra have been reduced with \textsc {iraf}\footnote{\textsc {iraf} is distributed by the National
Optical Astronomy Observatories}.  We used the overscan area of the Charge Coupled Device (CCD) for bias
subtraction. The data were flatfield corrected and optimally extracted (\citealt{1986PASP...98..609H}).
Wavelength calibration was done using lines from He, Ar \& Ne lamp spectra that were obtained during
daytime the day after the observations with the same instrument set--up, as is customary for VLT Service
mode observations. The rms scatter of the wavelength calibration was in the range of 0.05--0.08~\AA.

The extracted spectra were further reduced and analysed using the software package \textsc {molly}. We
corrected the wavelength calibration for potential shifts caused by flexure by cross correlating the spectra
over the wavelength ranges 9050--9160~\AA~and 9300--9500~\AA~with the first object spectrum. That part of the
spectrum is dominated by features from the night sky which should have the same wavelength in each spectrum.
Note that this does not account for uncertainties in the wavelength calibration caused by centroiding or star
tracking inaccuracies. We did this for both the spectra of Cir~X--1 and that of star 2 that was also in the
slit. Next, the observation times were corrected to the Heliocentric Julian Date time frame (using UTC times)
and we normalised and rebinned the spectra to a uniform velocity scale removing the Earth's velocity.
Cross--correlation of the spectra of star 2 with the first spectrum of that star over the range
8000--9000~\AA~shows that the remaining rms velocity differences are 2.5 km s$^{-1}$. 

\begin{figure} \includegraphics[angle=0,width=8cm,clip]{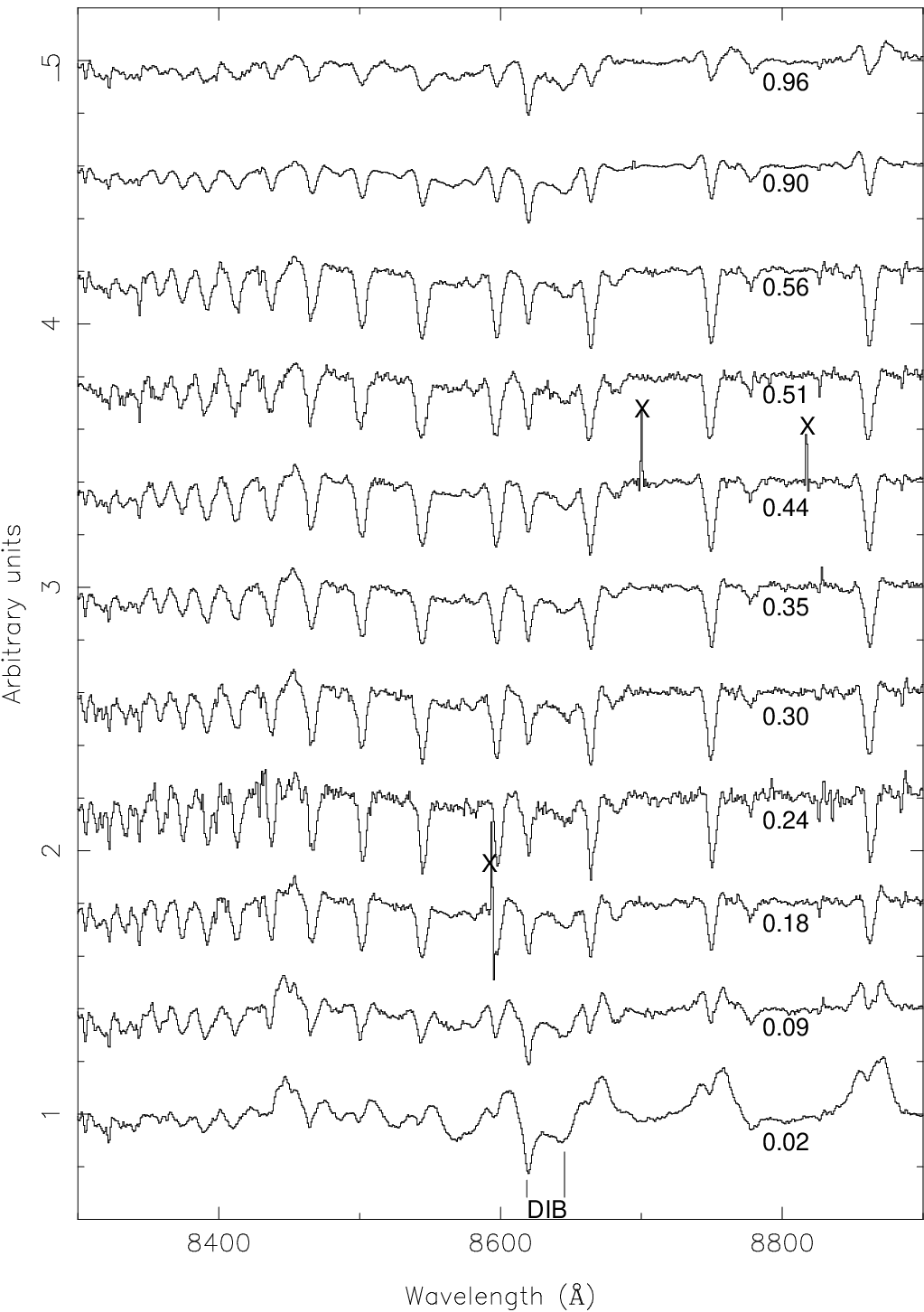} \caption{VLT/FORS2 spectra of
Cir~X--1 ranging from 8300--8900~\AA~for the different orbital phases as indicated on the right hand side in
the plot. The  Paschen lines are easily discernible. They change from absorption to strong emission lines
(with an absorption core) near phase zero. An arbitrary offset has been applied to the spectra for display
purposes. The X--es mark regions where cosmic ray hits have been incompletely corrected for. The absorption
line spectrum likely reveals the companion star. The line at 8619.5~\AA~is due to a diffuse interstellar
absorption band (DIB). The feature near  8650~\AA~is also due to DIB although an He~I line at 8648~\AA~found
in B stars may contribute to the broad nature of the line.  }

\label{fig:sample}
\end{figure}

In Fig.~\ref{fig:sample} we show the normalised spectra as a function of the orbital period. To fold the spectra
we used our best--fit epoch of periastron (T, see below) and an orbital period of ${\rm P_{orb}=16.54}$ days as
found by extrapolating the X--ray dip ephemeris of \citet{2004MNRAS.348..458C}. We binned the data in 15 phase
bins. However, none of the spectra fall in the phase range 0.60--0.85, hence only 11 phase bins are shown. The
most striking feature is the large change in the profile of the Paschen lines. At phase zero, large, broad
(possibly double peaked) emission lines are present. Superposed is an absorption line spectrum that could
originate in the companion star. This absorption line spectrum can be seen clearly at phases 0.2--0.6. Besides the
Paschen absorption lines that are often seen in B and A stars, small absorption lines often seen in supergiants
can be seen as well, e.g.~near 8777~\AA~(\citealt{1999A&AS..137..521M}; \citealt{2005A&A...434..949C}; compare the
Cir~X--1 spectrum with that of the supergiant system XTE~J1739--302 in \citealt{2006ApJ...638..982N}). This
feature is most likely due to He~I. 

\begin{figure} \includegraphics[angle=0,width=8cm,clip]{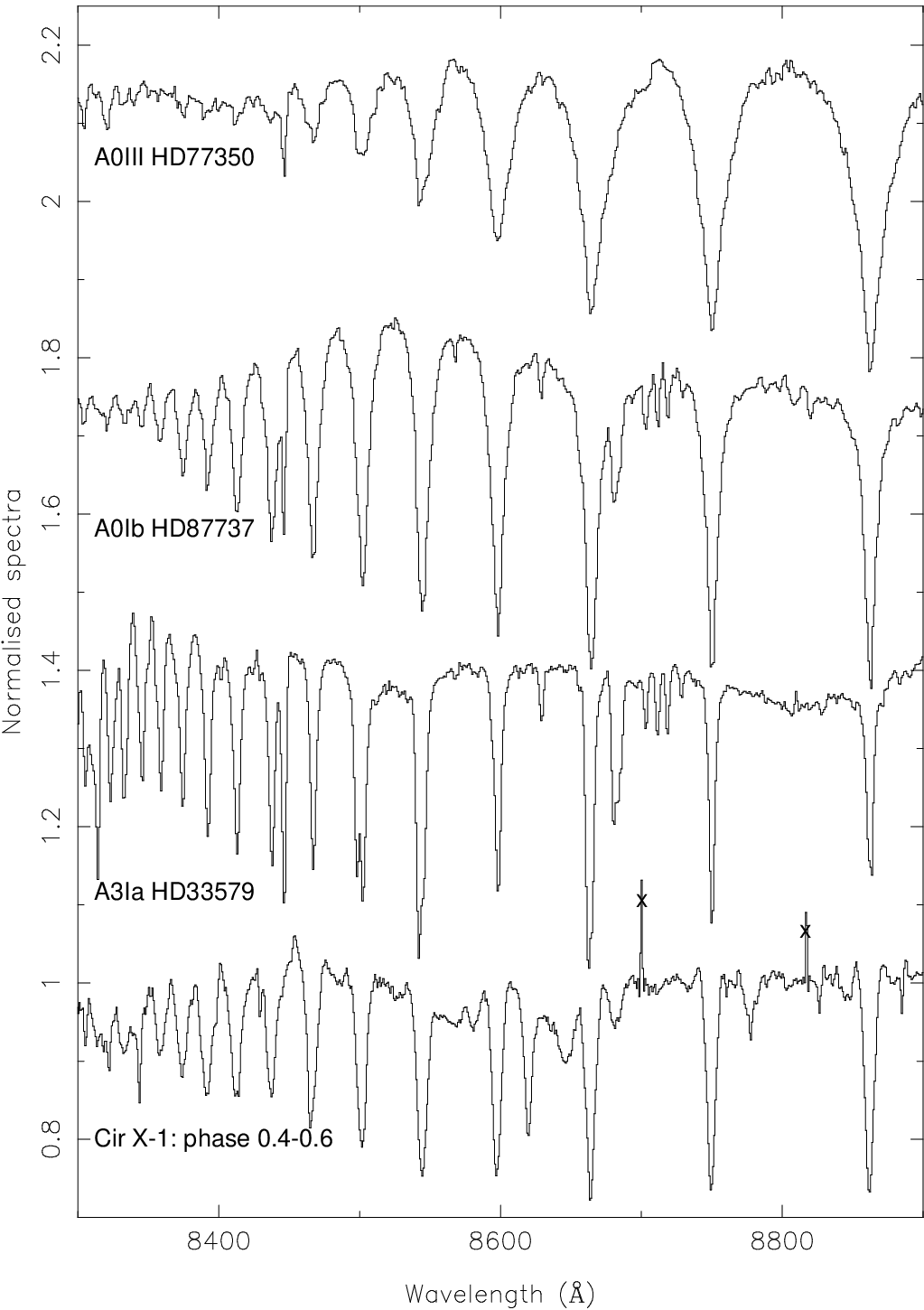}

  \caption{From {\it bottom} to {\it top}: Spectra ranging from 8300--8900~\AA~of Cir~X--1 (at $\sim$phase
  0.5; this work), the template star HD~33579 (spectral type A3Ia), HD~87737 (spectral type A0Ib) and
  HD~77350 (spectral type A0III; the latter three spectra are from \citealt{2003A&A...402..433L}). If the
  absorption line spectrum is from the companion star of Cir~X--1, the narrow Paschen lines show that the
  companion star must be a supergiant. The X--es mark regions where cosmic ray hits have been incompletely
  corrected for. }

\label{fig:comp}
\end{figure}

We investigated the behaviour of the equivalent width of the luminosity indicator Paschen--12 at 8750~\AA~as
a function of the orbital phase. At phase 0.4--0.6 the equivalent width of the absorption line is largest.
Assuming that the Paschen lines are formed in the companion star, the contamination of the accretion disc at
phase 0.4--0.6 is thus minimal but not necessarily zero. Hence, the measured equivalent width of 2~\AA~at
phase 0.4--0.6 is a lower limit. Using \citet{1994PASP..106..382D}, \citet{1995AJ....109.1379A},
\citet{1992ApJS...81..865S} and \citet{1999A&AS..137..521M} we find that the spectra at phase 0.4--0.6
resemble most that of late B/early A stars. As can be seen in Fig.~\ref{fig:comp}, the lines are very narrow
and from this we conclude that the star must be a supergiant. However, we cannot exclude that the absorption
line spectrum is from an accretion disc. Nevertheless, in order to test what spectral type would describe the
observed spectrum best under the assumption that the absorption lines are stellar, we obtained template
stellar spectra from \citet{2003A&A...402..433L}. We optimally subtracted the spectra of 9 stars (see
Table~\ref{tab:ref}) with spectral type ranging from B1Ib--F5Ia from the Cir~X--1 $\approx
8250-9300$~\AA~spectrum at phase 0.4--0.6. The optimal subtraction is performed on normalised spectra,
minimising the residuals using the following recipe: ${\rm f^{CirX-1}_\lambda=A+bf^{temp}_\lambda}$, where A
is the assumed accretion disc contribution and b is the fraction of light from the companion star. From the
residuals of the optimal subtraction it was clear that the Ia supergiants matched the narrow absorption lines
better than those of the Ib stars. Between the Ia supergiants we found that the spectra of those of spectral
type mid--B provided the best fit. The multiplicative factor in the optimal subtraction was consistent with
unity. However, systematic uncertainties prevent us from obtaining formal (reduced chi--squared,
$\chi_\nu^2$) fits. Since the template star spectra and the Cir~X--1 spectrum are not obtained with the same
telescope--instrument combination the instrumental broadening of the line profile is different. Furthermore,
the  broadening due to the rotational velocity of the stars is different for the different stars.

\begin{table}
\caption{Template stars from \citet{2003A&A...402..433L}.}
\label{tab:ref}
\begin{center}
Spectral type \& Henry Draper Identification \\
\begin{tabular}{cccc}
\hline
B1Ib & HD~091316 & B2Ia & HD~268623 \\
B5Ib & HD~164353 & B3Ia & HD~271163 \\
B7Iab& HD~268749 & B9Ia & HD~032034 \\
A0Ib & HD~087737 & A3Ia & HD~033579 \\
-- & --  & F5Ia & HD~269697 \\
\end{tabular}
\end{center}
\end{table}


\begin{figure*} \includegraphics[angle=-90,width=13cm,clip]{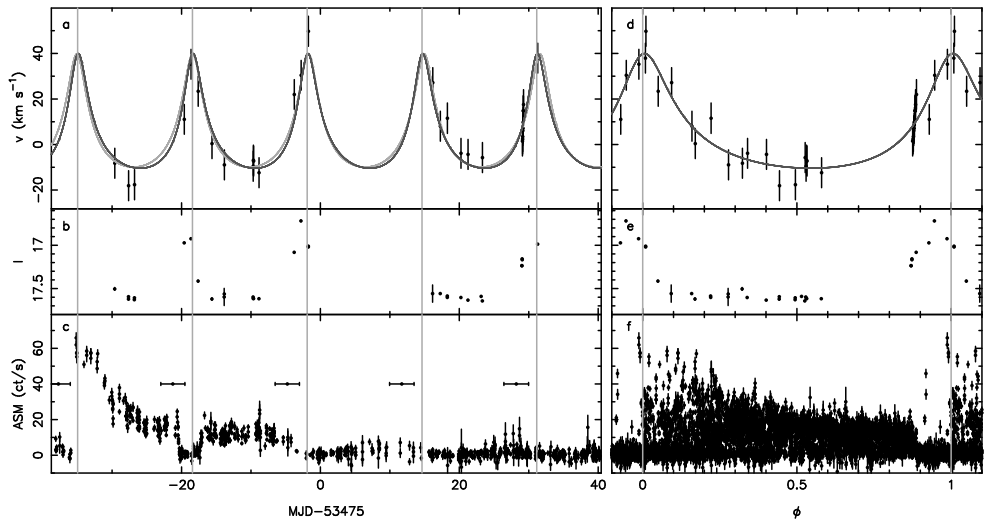}

  \caption{{\it Left panels:} (a) The radial velocity curve derived by cross--correlating the Cir~X--1 spectra
  over the range 8700--8900~\AA~(note that the velocities are relative to that of the template spectrum).
  Overplotted are the best--fit elliptical orbit treating the orbital period as a free parameter (light grey
  line) and that using the predicted orbital period from the X--ray dip ephemeris of Clarkson et al.~(2004;
  dark grey line, P$_{orb}=$16.53 days; Note that our best--fit orbital period is consistent with that derived
  from the ephemeris of Clarkson et al.~2004). (b) The evolution of the $I$--band magnitude as derived from
  psf--fitting on the $I$--band acquisition images. We have plotted the relative errors in the magnitudes, in
  addition there is the $\sim 0.2$ magnitude uncertainty in the zeropoint (i.e.~in the ordinate) (c) The
  contemporaneous X--ray flux as observed by the All Sky Monitor (ASM) on board the {\it Rossi} X--ray Timing
  Explorer. The points with horizontal error bars at an ASM count rate of 40 counts s$^{-1}$ indicate the
  predicted time of the X--ray dip according to the X--ray dip ephemeris of Clarkson et al.~(2004). {\it Right
  panels:} (d) The radial velocity curve folded on the X--ray dip orbital period of 16.54 days using the T
  from our best--fit to the radial velocity curve. (e) The phase folded $I$ band magnitude. (f) The phase
  folded but unbinned ASM lightcurve using data 10 cycles before and after MJD 53475 (dwell data from
  MJD~53140--53807). The X--ray dip phase and T from the radial velocity curve are consistent at the
  2~$\sigma$ level.}

\label{fig:radvel}
\end{figure*}

Next, we investigated the velocity of the absorption line spectrum as a function of the binary orbital
phase. Since we did not obtain early type supergiant template star spectra we cross correlate the spectra of
Cir~X--1 with that of one of the spectra. In the cross correlation it is vital that the continuum level has
been determined accurately. Furthermore, one has to avoid regions of the spectrum affected by skylines. In
order to avoid problems with the continuum normalisation and skylines we have cross--correlated the region
between 8700--8900~\AA. In Fig.~\ref{fig:radvel} we show the observed radial velocity curve. The solid line
in Fig.~\ref{fig:radvel} is the best--fit elliptical orbit (see Table~\ref{tab:radvel} and ~\ref{tab:ell}).
However, in the fit shown in Fig.~\ref{fig:radvel} we have increased the errors by 6.5 km s$^{-1}$ in order
to obtain a $\chi_\nu^2$ of 1.0. As mentioned above the rms residual velocities measured by cross
correlating the spectra of the nearby star with itself are 2.5 km s$^{-1}$. We used this 2.5 km s$^{-1}$ as
a measure of the amplitude of systematic effects. This systematic error dominates the statistical error of
the cross--correlation of the Cir~X--1 spectra. However, even when we take this uncertainty into account,
the formal $\chi^2_\nu\approx 6$ for the 25 degrees of freedom. Possible other systematic effects that are
not included are: (i) the effects of differences in X--ray heating on the absorption line spectrum. E.g.~it
can be seen comparing the {\it top and bottom left panel} of Fig.~\ref{fig:radvel} that the first three
velocity measurements are somewhat below the fit, these observations were performed when the X--ray flux was
still relatively high (ii) from Fig.~\ref{fig:sample} it can be seen that near phase zero strong Paschen
emission lines are present. These emission lines are redshifted with respect to their rest wavelengths and
since the Paschen absorption lines at those orbital phases are red shifted by a different amount, the
emission lines fill--in the absorption in an asymmetric way, possibly skewing the cross correlation
velocities (iii) slit centroiding and tracking errors can have introduced errors in the wavelength
calibration.

\begin{table}
\caption{Relative radial velocities of the counterpart of Cir~X--1 as displayed in Fig.~\ref{fig:radvel}a. To
convert these velocities to absolute velocities add the systemic velocity of $\sim -26\pm 3$ km s$^{-1}$.}
\label{tab:radvel}
\begin{center}
\begin{tabular}{cccc}
\hline
MJD & Orbital phase & Orbital phase & Radial velocity \\
   & ($P_{orb}=16.53$ d) & ($P_{orb}=16.68$ d) & (km s$^{-1}$)$^a$ \\
53445.3868 & 0.323 & 0.326  & -8.2$\pm$1.0\\
53447.3749 & 0.443 & 0.446  & -18.1$\pm$1.0\\
53448.2275 & 0.495 & 0.497  & -17.7$\pm$1.1\\
53455.3966 & 0.929 & 0.926  & 11.1$\pm$0.8\\
53456.3680 & 0.988 & 0.985  & 35.3$\pm$1.0\\
53457.4003 & 0.050 & 0.047  & 23.4$\pm$1.0\\
53459.3877 & 0.171 & 0.166  & 0.4$\pm$1.2\\
53461.1724 & 0.279 & 0.273  & -8.9$\pm$1.1\\
53465.3265 & 0.530 & 0.522  & -9.5$\pm$1.3\\
53465.3517 & 0.531 & 0.523  & -6.9$\pm$1.4\\
53465.3763 & 0.533 & 0.525  & -7.3$\pm$1.1\\
53466.1605 & 0.580 & 0.572  & -12.4$\pm$1.1\\
53471.2432 & 0.888 & 0.876  & 22.0$\pm$0.8\\
53472.2192 & 0.947 & 0.935  & 30.4$\pm$0.8\\
53473.2753 & 0.011 & 0.998  & 49.7$\pm$1.1\\
53491.1779 & 0.094 & 0.072  & 27.2$\pm$1.0\\
53492.2715 & 0.160 & 0.137  & 8.1$\pm$1.1\\
53493.2927 & 0.222 & 0.198  & 11.5$\pm$1.6\\
53495.2467 & 0.340 & 0.316  & -3.9$\pm$0.9\\
53496.2661 & 0.402 & 0.377  & -4.3$\pm$1.0\\
53498.3358 & 0.527 & 0.501  & -5.7$\pm$1.2\\
53504.0955 & 0.875 & 0.846  & 1.5$\pm$0.9\\
53504.1153 & 0.877 & 0.847  & 2.4$\pm$0.8\\
53504.1351 & 0.878 & 0.848  & 3.5$\pm$0.9\\
53504.1549 & 0.879 & 0.850  & 6.1$\pm$0.8\\
53504.1748 & 0.880 & 0.851  & 7.1$\pm$0.8\\
53504.1946 & 0.881 & 0.852  & 9.8$\pm$0.9\\
53504.2144 & 0.883 & 0.853  & 14.9$\pm$0.9\\
53504.2343 & 0.884 & 0.854  & 14.7$\pm$1.0\\
53504.2541 & 0.885 & 0.856  & 17.5$\pm$1.0\\
53506.3040 & 0.009 & 0.978  & 38.0$\pm$0.8\\
\end{tabular}
\end{center}
{\footnotesize$^a$ The error has to be increased by 6.5 to obtain a reduced $\chi^2$ of $\approx$1 in
the fit (see text).}\\
\end{table}

Since we cross correlated the spectra with a spectrum of Cir~X--1 itself the systemic velocity that is derived
from the elliptical orbit fit (3.7$\pm$1.3 km s$^{-1}$) is not the true systemic velocity. In fact, as one
would expect, within 3~$\sigma$ it is consistent with 0. The systemic velocity is $\sim -26\pm 3$ km s$^{-1}$.
We derived this by comparing the best--fit central wavelength of the Gaussian absorption lines with rest
wavelengths of 8467.25, 8598.4, 8750.5 and 8862.8~\AA. The Gaussians were fitted to the spectrum that was taken
as our template spectrum in the cross--correlation (we have excluded the three Paschen lines that are possibly
blended with absorption lines of the Ca~II triplet). Hence, in order to convert the relative velocities in
Fig.~\ref{fig:radvel} and Table~\ref{tab:radvel} to absolute velocities one has to add the systemic velocity.

\begin{table} \caption{The best--fit orbital parameters. We provide both the solution for the fit with the
orbital period as a free fit--parameter and that for where it was fixed to the orbital period extrapolated
from the ephemeris of \citet{2004MNRAS.348..458C}. }

\label{tab:ell}
\begin{center}
\begin{tabular}{lcc}
\hline
P (days)& 16.68$\pm$0.15 & 16.53 (fixed)\\
K (km s$^{-1}$) & 25$\pm$2& 25$\pm$2\\
e & 0.45$\pm$0.07& 0.47$\pm$0.06\\
$\omega$ (deg) & 2$\pm$12 & -7$\pm$11\\
T (+53,475.0 days; MJD) & -1.7$\pm0.4$& -1.9$\pm0.4$\\
a $\sin$ {\it i} (lightseconds) & 16.9$\pm$1.2 & 16.8$\pm$2.0\\
f(m) (M$_\odot$) & 0.019$\pm$0.007 & 0.019$\pm$0.007\\

\end{tabular}
\end{center}
\end{table}

We have used the 32 $I$--band images that were obtained to acquire the source to construct a lightcurve.
The images were corrected for bias using the values from the overscan regions and flatfielded using sky
flats taken within one or two days from the science images. To determine instrumental magnitudes of
Cir~X--1 and stars in its relative crowded vicinity, we used the point--spread--function (psf) fitting
routines from  \textsc {DAOPHOT~II} (\citealt{1987PASP...99..191S}), running inside \textsc {midas}. The
instrumental magnitudes were placed in a common photometric system by removing small magnitude differences
between the different images, due to e.g.\ differences in the psf and the exposure times, by matching stars
between these images and comparing their magnitudes. This common system was then calibrated using
observations of standard star fields (PG\,0942$-$029 and L110; using the calibrated magnitudes by
\citealt{2000PASP..112..925S}), which were imaged during photometric nights (March 27, April 11 and 12,
2005). As Cir~X--1 is only imaged in the $I$--band, we have not determined colour terms. The uncertainty in
the zeropoint is estimated to be about 0.2\,mag. The derived evolution of the $I$--band magnitude as a
function of time and orbital phase is shown in the {\it middle panels} of Fig.~\ref{fig:radvel}. It is
clear that the $I$--band magnitude is constant except for a brief period that coincides with the X--ray dip
and periastron passage.

\section{Discussion}

We have obtained VLT phase resolved optical $I$--band spectra and images of the X--ray binary Cir~X--1. The
observed X--ray flux was low compared to that observed during other earlier spectroscopic observations
(e.g.~\citealt{1999MNRAS.308..415J}; \citealt{2001MNRAS.328.1193J}; \citealt{2003A&A...400..655C}). If this lower
X--ray flux corresponds to a lower intrinsic X--ray luminosity the optical light from the accretion disc thought to
be caused in large by reprocessing of the X--ray luminosity might be lower as well. Hence, the putative companion
star might be more readily observable now compared with high X--ray flux episodes. The spectrum we observed varies
strongly as a function of orbital phase. Strong emission lines appear near periastron passage, filling--in the
absorption line spectrum observable at other orbital phases. Cross--correlation of the absorption features in the
8700--8900~\AA~range gives a radial velocity curve with an orbital eccentricity of $e=0.45$, $a\,\sin\,i=16.9$
lightseconds and a mass function of 0.019 M$_\odot$. The question is; is this the radial velocity curve of the
companion star, is it associated with the compact object/neutron star, or are the absorption lines caused by a
circumbinary disc?

The detected absorption line spectrum is consistent with a stellar spectrum if that star is a B5--A0 supergiant. This
provides evidence for a supergiant companion star in Cir~X--1. Similarly, \citet{2003A&A...400..655C} found emission
lines in the near--infrared spectrum of Cir~X--1 that often occur in mid--B supergiants. Those results are consistent
with our findings (but \citealt{2003A&A...400..655C} favoured a scenario where the emission features arise in the
accretion disc/flow.) Previously, \citet{1980A&A....87..292M} proposed that the companion star of Cir~X--1 is an early
type supergiant. However, the then known counterpart was later resolved into three stars reducing the magnitude
associated with the companion star (\citealt{1992A&A...260L...7M}) which led people to discard the supergiant model for
Cir~X--1. On the other hand, an interstellar extinction of A$_V=5$ was used. This value was derived on
the basis of the lowest N$_H$ measured in the seventies. That N$_H$ was determined from an X--ray spectrum obtained from
Aerobee rocket data. No error was given on N$_H$ (\citealt{1971ApJ...169L..23M}). More recent N$_H$ measurements from
X--ray spectral fits with more sensitive satellites with a good soft response such as ROSAT, ASCA and {\it Chandra} give
an N$_H$ in the range 1.6--2.2$\times10^{22}$ cm$^{-2}$ (\citealt{1995A&A...293..889P}; \citealt{1996MNRAS.283.1071B};
\citealt{1999ApJ...511..304S}; \citealt{2002ApJ...572..971S}). However, \citet{2005ApJ...619..503I} again found a
significant lower value for N$_H$ modelling {\it Beppo}--SAX data. As we will show below there are three independant
measurements that are consistent with the higher N$_H$ for Cir~X--1. First of all, as mentioned by
\citet{1996MNRAS.283.1071B}, a significantly lower value for N$_H$ would not be compatible with the strength of the dust
scattering halo in Cir~X--1 as found by \citet{1995A&A...293..889P}. Secondly, the distance that would be derived from
the low N$_H$ value is $\sim$4 kpc (\citealt{2005ApJ...619..503I}). This distance is too low to explain the type I X--ray
bursts, those require a lower limit to the distance of d=7.8--10.5 kpc (\citealt{2004MNRAS.354..355J}). Finally, the
equivalent width of the diffuse interstellar band at 8620~\AA~that we find in the optical spectrum is correlated with
$E_{(B-V)}$ (\citealt{2000msl..work..179M}). This provides an A$_V=12.2$ for Cir~X--1 (taking R=3.1), consistent with the
higher value derived from N$_H$ which using the conversion of N$_H$ to A$_V$ from \citet{1995A&A...293..889P} gives
$9<{\rm A}_V<12$. For these reasons and since the modelling of the X--ray spectrum is model dependent as mentioned in
\citet{2005ApJ...619..503I} we think that the higher value for N$_H$ and hence A$_V$ is more likely to be right.

From the distance to Cir~X--1 from the type I X--ray bursts (d=7.8--10.5 kpc), A$_V\approx 9-12$, and $I$=17.6, one gets an
absolute magnitude $-4.9<{\rm M}_V<-2.5$ (taking into account that $V-I$ is nearly 0 for mid--B/early A type stars, and
using the relative extinction from  \citealt{scfida1998} A$_I=0.6\times$A$_V$). An absolute magnitude of -2.5 would be
too low and even -4.9 is on the low side for standard B5--A0 supergiants (\citealt{1974MNRAS.166..203B}). However, considering the preceding binary
evolution involving significant mass transfer from the neutron star progenitor to the initially less massive companion
star and the current X--ray heating especially at periastron the companion star of Cir~X--1 is not likely to be a
standard supergiant.

\begin{figure} \includegraphics[angle=0,width=7cm,clip]{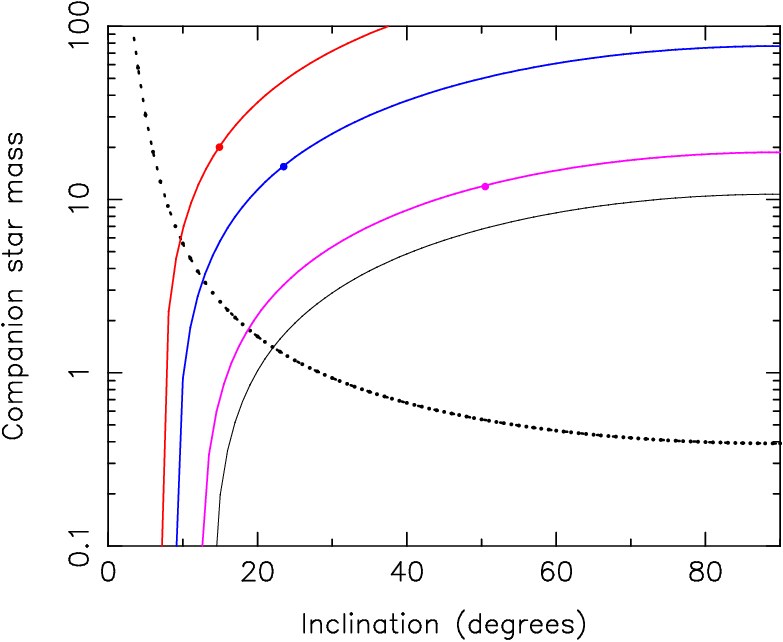}

  \caption{The mass of the companion star as a function of the orbital inclination with respect to the
line--of--sight. The drawn lines are derived assuming that the compact object is from left to right, a 10, 5, 2,
or a 1.4 M$_\odot$ compact object. The dots are solutions if imposing the radius of the companion star to be
smaller than or equal to the periastron distance $a\,(1-e)$ (we took the least constraining case of B5I, the
radius of a A0I is somewhat larger hence the dots would all move to lower inclinations). For each of the solid
line -- dot combination the allowed parameter space is that on the solid line and below the dot. This yields the
following constraints on the companion star mass for a 10, 5, 2 and 1.4 M$_\odot$ compact object; ${\rm M_{comp}
\approxlt 18.9\,M_\odot}$, $\approxlt 15\,{\rm M_\odot}$, $\approxlt 11.9\,{\rm M_\odot},$ and $\approxlt
10.4\,{\rm M_\odot}$, respectively. The inclination is constrained to be $ 7.1^\circ\approxlt i \approxlt
14.9^\circ$, $8.9^\circ\approxlt i \approxlt 23.5^\circ$, $12.1^\circ\approxlt i \approxlt 50^\circ$ and $i
\approxgt 13.7^\circ$ for a 10, 5, 2, and 1.4~M$_\odot$ compact object, respectively. Taking an AOI star instead
of a B5I would change the inclination and companion star mass constraint to $ 7.1^\circ\approxlt i \approxlt
10^\circ, 7.3$ M$_\odot$ for the 10 M$_\odot$ compact object. Under the assumption that the measured $a\, \sin\,i$
is not $a_{opt}\, \sin\, i$ but $a_{NS}\, \sin\, i$, the constraint on the companion star mass as a function of
inclination is given by the dotted line (taking ${\rm M_{NS}=1.4\,M_\odot}$).}

\label{fig:massinc}
\end{figure}

Let us investigate the consequences of the assumption that the absorption line spectrum and the radial velocity curve
track the companion star of Cir~X--1. Kepler's third law, the measured a$_{opt}\, \sin$ {\it i} and P$_{orb}$ define
a relation between the inclination of the orbit with respect to the line--of--sight and the mass of the companion
star (the solid lines in Fig.~\ref{fig:massinc}). From the spectral classification B5--A0I we find $2.1\approxlt
\log\,g\, \approxlt 2.4$ (\citealt{2000asqu.book.....C}). This gives
$\frac{R_{comp}}{R_\odot}=C\sqrt\frac{M_{comp}}{M_\odot}$ (where $C$ is in the range of 10--15). To avoid the neutron
star going through the companion star at periastron, we constrain the radius of the companion star to be smaller than
or equal to the periastron distance  $a\,(1-e)$, with $a\,\sin\,i={\rm \frac{M_X+M_2}{M_X}} a_{opt}\,\sin\,i$ and
$a_{opt}\,\sin\,i =7.28$ R$_\odot$. From Fig.~\ref{fig:massinc} this yields ${\rm M_{comp} \approxlt 18.9\,M_\odot}$,
$\approxlt 15\,{\rm M_\odot}$, $\approxlt 11.9\,{\rm M_\odot},$ and $\approxlt 10.4\,{\rm M_\odot}$ for
M$_{comp}=10$, 5, 2 and 1.4 M$_\odot$, respectively. The inclination is constrained to be $ 7.1^\circ\approxlt i
\approxlt 14.9^\circ$, $8.9^\circ\approxlt i \approxlt 23.5^\circ$, $12.1^\circ\approxlt i \approxlt 50^\circ$ and $i
\approxgt 13.7^\circ$ for a 10, 5, 2, and 1.4~M$_\odot$ compact object, respectively. A normal B5--A0 supergiant has
a mass of $\sim$10 M$_\odot$ which would hence fit--in with a neutron star compact object of Cir~X--1. On the other
hand it is conceivable that if the companion star of Cir~X--1 has indeed spectral type B5--A0I it is no ordinary star
due to the preceding evolution (as mentioned above). 

The observed type~I X--ray bursts (\citealt{1986MNRAS.219..871T}) are evidence of a neutron star nature of the
compact object. Furthermore, as mentioned in the Introduction the neutron star has to have a magnetic field
$\approxlt 10^{12}$ Gauss. Also the presence of a strong radio jet is not compatible with a high magnetic field
neutron star. These findings seem to be at odds with a supergiant (thus young) companion star. We conclude that if
the presence of a supergiant companion star in Cir~X--1 is confirmed the neutron star magnetic field either
decayed quickly to below $\approxlt 10^{12}$ Gauss or the neutron star was born with such a low field.

Assuming that a neutron star kick at birth did not give the binary system a large systemic velocity, one can
calculate the radial velocity for Local Standards of Rest along the direction of Cir~X--1 (see for instance
\citealt{2001ApJ...555..364B}) to find that the systemic radial velocity of $\sim -26\pm 3$ km s$^{-1}$ at the
location of Cir~X-1 gives a distance of either $\sim 1.6$ kpc or $\sim$11.8 kpc. A distance of 11.8 kpc is close to
the distance derived from the observed type I X--ray bursts (\citealt{1986MNRAS.221P..27T}; 7.8--10.5 kpc
\citealt{2004MNRAS.354..355J}).

The optical I--band lightcurve (see Fig.~\ref{fig:radvel}e) shows a clear brightening near periastron. This is similar in shape to
the lightcurves published by \citet{1992A&A...260L...7M} and \citet{1994MNRAS.268..742G}, although one has to bear in mind that
those lightcurves were obtained when the observed X--ray luminosity of the source was much higher and that the data was phase folded
using different ephemerides. The increase in I--band light corresponds to the phase where Paschen emission lines start to become
apparent in the spectrum. Hence, it is probable that the enhanced mass transfer rate near periastron is responsible for both
effects. For instance, the emission lines could be formed in the accretion stream from the companion star to the compact object and
the enhanced I--band emission could be both due to these emission lines as well as due to enhanced continuum emission related to the
stream impact site. 

In principle it cannot be ruled out that the absorption line spectrum originates in the accretion disc.
E.g.~the $I$--band spectrum of the high inclination (accretion disc corona) source 2S~0921--630 shows Paschen
absorption lines at orbital phases near 0.9 that could well be caused by the line of sight passing through the
accretion disc rim (\citealt{2005MNRAS.356..621J}). Those Paschen absorption lines are not present in
the K1III spectral type that has been derived for the companion star in 2S~0921--630 (although X--ray heating
effects may result in different spectral types being observed at different orbital phases, cf.~the observed
spectral type in Cyg~X--2 changes from A5--F2 \citealt{1979ApJ...231..539C}). However, in radio observations
of Cir~X--1 superluminal motion has been observed which limits the inclination of the jet axis to the line of
sight to $i<5^\circ$ (\citealt{2004Natur.427..222F}; the limit $i<5^\circ$ has been derived assuming a
distance of 6.5 kpc, if Cir~X--1 is indeed further away as is indicated by the burst properties, then the
limit on $i$ is more stringent still). If our line--of--sight also goes through the accretion disc rim (at all
orbital phases) it implies that the jet--axis is nearly in the plane of the orbit. Hence, it implies that the
jet ploughs through the accretion disc if it originates close to the compact object.

Nevertheless, if we assume that the $a\, \sin\,i$ we measured is associated with the binary motion of the
accretion disc/neutron star we can derive a limit on the mass of the companion star assuming the neutron
star mass is 1.4 M$_\odot$. This limit is a function of the binary inclination and is given by the dotted
line in Fig.~\ref{fig:massinc}. From the discussion above it would seem that the companion star mass is
$\approx$0.4 M$_\odot$ since the binary inclination must be high in this scenario. A star of such a mass
cannot have evolved off the main--sequence in a Hubble time unless the star was more massive initially (at
least 0.8 M$_\odot$) and more than 0.4 M$_\odot$ has been transfered already. However, it is unclear whether
such an amount of mass can have been transfered while the orbital eccentricity is still 0.45.

Finally, in this paragraph we consider other possible formation scenarios for the absorption line spectrum. 
E.g.~in the peculiar X--ray binary SS~433 evidence for a circumbinary disc has been found
(\citealt{2001ApJ...562L..79B}), perhaps the absorption lines are formed in such a disc. However,
\citet{1988AJ.....96..242F} have obtained I--band spectra of SS~433 and their spectra are markedly different
from those that we observe from Cir~X--1. In SS~433 the Paschen lines are double peaked and in emission
whereas we find them to be in absorption except near periastron passage. On the other hand the inclinations
at which we observe SS~433 and Cir~X--1 are different. In this respect it is interesting to note that
according to \citet{2006astro.ph..7612T} we will be observing the system through the material that is swept
up by the approaching radio jet. However, it is difficult to imagine that that material is dense enough and
has a velocity low enough to  cause the narrow absorption lines. Finally, in all these scenario's the
velocity changes of the absorption lines with orbital phase are difficult to explain.

A potential way to determine whether the measured radial velocity curve is associated with the companion star is
to obtain a high resolution spectrum. Such a spectrum could reveal narrow spectral features and allow for a better
spectral classification e.g.~the observed absorption feature near 8685~\AA~consists of a blend of several narrow
lines if the spectrum is from the companion star (\citealt{1999A&AS..137..521M}) and if the companion star is an
A--star the Ca~II absorption triplet can be separated from the Paschen lines. Furthermore, the rotational
broadening of the spectral lines could be measured. Finally, due to precession of the orbit (i.e.~apsidal motion),
the anomalous period as measured by the time between periastron passages is different from the period that would
be measured by eclipse timing (i.e.~the sidereal period). Hence, if the X--ray dips in Cir~X--1 are due to
(grazing) eclipses the sidereal period will over time differ from the anomalous period.

\section*{Acknowledgments}  

\noindent The authors are grateful to the referee, Dr.~D.~Gies, for his comments that helped improve the
Paper. We would like to thank the Director of ESO for approving these DDT observations. The use of the
spectral analysis software package \textsc {molly} written by prof.~Tom Marsh is acknowledged.  PGJ
acknowledges support from NASA grants NNG05GN20G and NNG05GN27G. PGJ, GN and CGB acknowledge support from
the Netherlands Organisation for Scientific Research.

\end{document}